# Frequency-dependent photothermal measurement of thermal diffusivity for opaque and non-opaque materials; Application to crystals of TIPS-pentacene

**Maryam Shahi and J.W. Brill**
Department of Physics and Astronomy
University of Kentucky, Lexington, KY 40506-0055

**Abstract**

We propose the use of a frequency-dependent photothermal measurement as a complement to light-flash, i.e. time-dependent, measurements to determine the through-plane thermal diffusivity of small, thin samples, e.g. semiconducting polymers and small organic molecule crystals. The analysis is extended from its previous use with some opaque conducting polymers to materials with finite absorption coefficients, such as crystals of 6,13-bis(triisopropylsilylethynyl pentacene ("TIPS-pentacene"). Taking into account the finite absorption coefficients of the latter gives a value of diffusivity, $D \approx 0.10$ mm$^2$/s, much smaller than previously estimated and more consistent with its expected value. We also briefly discuss the effects of coating samples for the measurement to improve their optical properties.

## I. Introduction

As the variety of applications of organic semiconductors grows, it becomes increasingly important to determine values of their room temperature thermal conductivities. Because it is often difficult to apply contacts to the sample with sufficiently small interface thermal resistances, photothermal techniques, in which the sample is heated with light and the resulting thermal radiation used to determine the temperature change, are popular.[1,2] In particular, commercial light-flash apparatuses, in which an intense light pulse is used and the time dependence of the thermal radiation is measured are common; the characteristic time of the sample is proportional to $d^2/D$, where d is the thickness of the sample and the thermal diffusivity $D = \kappa/c\rho$, where $\kappa$ is the (through-plane) thermal conductivity, c the specific heat, and $\rho$ the mass density.[1,3] However, because of the time resolution of the instrument (typically > 0.1 ms), it is difficult to measure samples thinner than d ~ 100 μm, often a problem for new semiconducting polymers. In addition, light-flash techniques typically require samples with areas > 20 mm$^2$, difficult to achieve for single crystals of small molecule semiconductors such as "TIPS-pentacene" (6,13-bis(triisopropylsilylethynyl pentacene).[4,5]

We recently reported a simplified photothermal technique,[6] derived from Reference [2], in which the frequency dependence of the thermal radiation, when the incident light is chopped at a variable frequency, is measured. Working in the frequency domain allows the use of much less intense (and less expensive) light sources, and also allows one to measure samples at least one order of magnitude thinner and smaller area than light-flash techniques. In this paper, we include a more detailed description of the technique and analysis, which we also extend for use on materials which are not optically opaque. As an example, we show how our previous analysis for TIPS-pentacene,[6] which ignored the semi-transparency of the material in the infrared, led to a huge over-estimate of its interlayer thermal diffusivity.

## II. Experimental Technique and Analysis for Opaque Samples

The inset in Figure 1b shows a schematic of the apparatus.[6] The sample is glued to an aperture which is placed inside the dewar of a liquid nitrogen cooled mercury-cadmium-telluride (MCT) photoconducting detector (with net sensitivity ~ 0.4 V/µW), < 1 cm away from the detector. For a light source, we used either a mechanically chopped (f = 0.5 Hz – 2 kHz) quartz-halogen lamp, whose light was fed to the window of the dewar through a fiber-optic bundle, or a 447 nm, 1 W diode laser, whose light could either be mechanically or electronically (maximum f = 500 Hz) chopped; when using the laser, a ground glass diffusing plate is placed in front of the sample. When needed to attenuate incident light (that either passes through or around the sample), a 10 µm long-wave pass filter (LPF) is placed between the sample and detector. (Although the MCT responsivity peaks for mid-infrared wavelengths, it does have finite response for near IR and even visible wavelengths.) Silvered glass tubes are placed between the filter and detector and between the window and sample to maximize the detected and incident light intensities. The oscillating detector signal at the chopping frequency is measured with a 2-phase lock-in amplifier and normalized to the frequency dependence of the detector preamplifier. Our setup has the simplifying advantage over those of other reported ac-photothermal setups[2,7] in that the sample is in the same vacuum as the detector, eliminating the need for focusing mirrors and a window between the sample and detector.

If the sample is opaque to both the incident and thermal radiation, so that light is absorbed wholly on the front surface and the emitted light comes only from the back surface, the expected complex signal $V_{ac}$ is given by:[2]

$$f V_{ac} = f(V_X + iV_Y) = -A \chi / \Psi \quad (1a)$$
$$\Psi = (1+i) [\sinh \chi \cos\chi + i (\cosh\chi \sin\chi)] \quad (1b)$$
$$\chi \equiv (4.743 \, f/f_2)^{1/2} \equiv d \, (\pi f/D)^{1/2} \quad (1c)$$

The characteristic frequency $f_2 \equiv 1/(2\pi\tau_2)$, where $\tau_2$ is the conventional ac-calorimetric "internal" thermal time constant" describing heat flow through the sample.[8] The magnitude A is proportional to the absorbed light intensity and inversely proportional to the sample's heat capacity;[2] the negative sign reflects the fact that we set the lock-in so that the phase of the incident light = 180°. This choice of phase conveniently makes the in-phase ($V_X$) and quadrature ($V_Y$) signals positive at the lowest measured frequencies, as shown by the solid curves in the inset to Figure 1a, in which the in-phase and quadrature signals (times f) are plotted as functions of $f/f_2$. The assumptions made in deriving Eq. (1) are that the lateral dimensions of the sample are much larger than the thickness and that the sample is uniformly illuminated so that heat flow in the sample can be treated as one-dimensional[2,8] and that the chopping frequency f >> $1/(2\pi\tau_1)$, where $\tau_1$ is the "external" thermal time constant with which the sample comes to equilibrium with its surroundings,[8] typically greater than 1 second for our samples. $|V_{ac}| \propto 1/f$ for $f_1 << f << f_2$ and goes to zero more quickly as f exceeds $f_2$. (A typical indication of non-one dimensional heat flow due to finite thickness would be a reduction in $fV_Y$ at low frequencies,[9] while an indication of small $\tau_1$ would be a low-frequency reduction in $fV_X$.[8])

In Reference [6], we reported on measurements of opaque films of "NFC:PEDOT", cellulose nanofibrils coated with the conducting polymer blend, PEDOT:PSS [poly(3,4-ethylene-dioxythiophene):poly(styrene-sulfonate)], measured with the quartz-halogen light source. For those measurements, we did not yet have the long-wave pass filter; since any incident light which leaks through or around the samples adds a (negative, with our sign convention) term proportional to frequency to $fV_X$, we only reported on the quadrature signal. Figure 1a shows the measured frequency dependence of both the in-phase and quadrature signals from one of these samples with d = 62 µm and area < 10 mm$^2$, with fits to Eq. (1) with a negative "leaked light" signal added to the in-phase component. The fit has four parameters, the magnitude of the thermal signal, the magnitude of the leaked light, $f_2$, and a small phase error (typically a few degrees) in setting the phase of the lock-in, since the phase shift depends on the alignment of the optical system. Also shown are results on a second sample, of the same thickness, measured with the long-wave pass filter in-place and zero leaked-light assumed in the fit. The values of $f_2$ (119 Hz) for both samples are equal within their uncertainties (2 %) and correspond to a diffusivity value D = 0.30 mm$^2$/s.

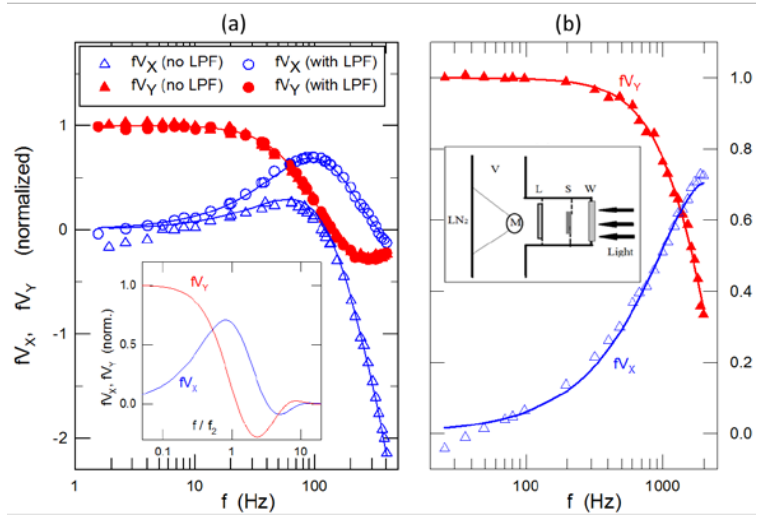

**Figure 1.** (color online) Frequency dependence of the frequency times in-phase ($V_X$) and quadrature ($V_Y$) signals for a) 62 µm thick samples of NFC:PEDOT and b) 10 µm thick sample of PEDOT:PSS (from Ref. [10]). The data has been normalized to the quadrature signal at low-frequency. Solid curves show fits of the data to Eq. (1). For NFC:PEDOT, results both with and without a long-wave pass filter are shown. Inset (a) shows the theoretical frequency dependence (Eq. 1). Inset (b) shows a schematic of the apparatus, with M = MCT detector, L = LPF, S = sample, V = vacuum space, W = glass vacuum window (reprinted from Ref. [6]).

The utility of the technique for very thin samples is shown in Figure 1b, for which we show the experimental results for a d = (10 ± 1) µm free-standing (also opaque) sample of PEDOT:PSS,[10] again measured with the quartz-halogen light source and assuming zero leaked-light. The fitted value of $f_2$ = (2.61 ± 0.09) kHz, corresponding to D = (0.17 ± 0.04) mm$^2$/s, where most of the uncertainty comes from that of the thickness. While our present apparatus is limited to frequencies below 2 kHz, this is not an intrinsic limitation, and even thinner samples could be measured, e.g. with a higher frequency electronically chopped laser.

A common practice in light flash analysis is to coat a sample which has "non-ideal" optical properties with opaque films, such as colloidal graphite.[11] Non-ideal properties include having low emissivity or low absorbance for thermal radiation and/or high reflectance or low absorbance for incident light. We will discuss the extension of our technique for low-absorbance samples in Section III. To investigate the effect of coating a sample with high reflectivity and low emissivity, we investigated a copper sample with d = (353 ± 8) μm using the diode laser, with the results shown in Figure 2. For the uncoated sample, the data was well fit with $f_2$ = (1277 ± 16) Hz, corresponding to D = (105 ± 6) mm$^2$/s, consistent with published results (111 mm$^2$/s).[11,12] Evaporating a ~ 100 nm PbS film on the front (incident) surface to decrease the reflectivity and increase the magnitude of the absorbed light approximately doubled the signal but did not change $f_2$, as shown in the figure. We then removed the PbS and deposited a graphite film (between 5 -10 μm) on the front surface. As shown in the figure, the characteristic frequency decreased by almost an order magnitude. We also investigated covering the back surface with a similar graphite film (removing the film from the front surface); in this case, as shown in the figure, the signal increased by an order of magnitude and the response time was faster than for the front film, as expected, but still a few times slower than that of the uncoated sample.

These results illustrate that common graphite films can have time constants on the order of a millisecond and are generally not practical for the thin organic samples for which our technique is designed. (Light-flash techniques generally apply films for samples with characteristic times greater than several ms.[11]) Appropriate evaporated films can be chosen to enhance the signal, but for front-surface films, care should be taken that the film is uniform and sticks well to the surface.

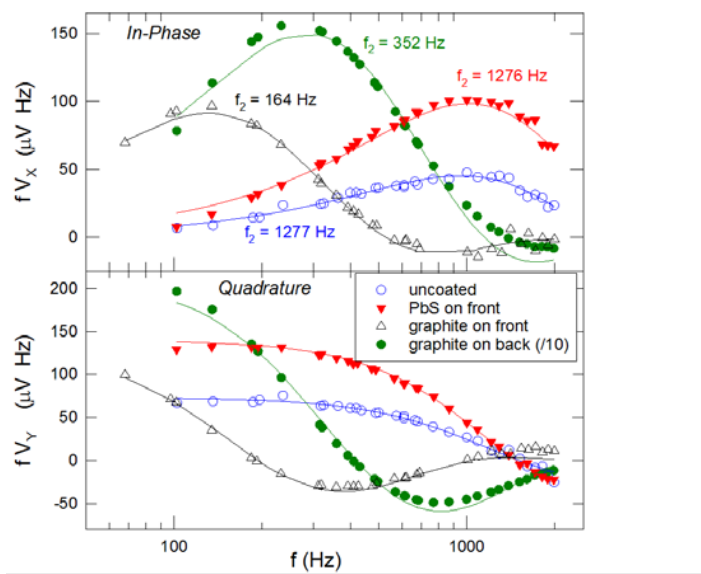

**Figure 2.** (color online) Frequency dependence of signals of a 353 μm thick copper sample, with and without coatings, as indicated. The curves show fits to Eq. 1, with the fitted values of $f_2$ indicated. The (non-normalized) absolute values of the detector signals are given to show the effects of coating. (Note that the vertical scales for the in-phase and quadrature responses are different and that the signal for the sample with graphite on back is 10 times larger than shown.)

### III. Analysis for non-Opaque Samples: TIPS-pentacene

TIPS-pentacene[4,5] is a model small molecule organic semiconductor with a layered, brick-work structure; some of the recent work on the electronic and structural properties of this material are listed in Ref. [13]. In Ref. [6], we reported on photothermal measurements of its through-plane thermal diffusivity. In that work, we used the incandescent quartz-halogen light source, concentrated on the quadrature signal because there was a large amount of leaked light (which we assumed leaked around the irregularly shaped crystals), but assumed that the sample was sufficiently opaque. This assumption led us to a very large value of the interlayer diffusivity and thermal conductivity, values an order of magnitude larger than generally found in organic materials, which we tentatively associated with interactions between rotations of the TIPS side-groups which extend between the layers of the crystal. However, subsequent measurements on sublimed thin films of TIPS-pentacene deposited on substrates[14] gave a value for the interlayer thermal conductivity two orders of magnitude smaller than the value we calculated for crystals in Ref. [6]. While the thin films are not fully ordered, it seemed unlikely that the disorder could account for the two order of magnitude reduction in thermal conductivity, further motivating us to reconsider our previous analysis.

While fairly opaque for visible light, TIPS-pn is in fact quite transmitting throughout the infrared, as shown in Figure 3. If the absorption length for incoming light ($1/\alpha$) is not much smaller than the sample thickness, then incident light will also heat the center of the sample, speeding up the thermal response on the back surface. Similarly, if the absorption length of thermal radiation ($1/\beta$) is not much less than d, radiation from the interior can reach the detector, so heat does not need to diffuse through the whole thickness to contribute to the signal. Consequently, Eqtn. (1a) must be generalized:[2]

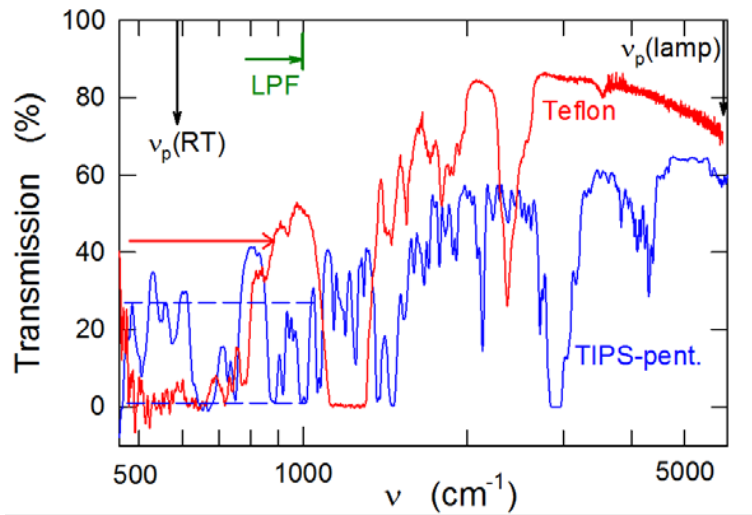

**Figure 3**. (color online) Infrared transmission spectra of a d = 190 μm thick crystal of TIPS-pentacene and a d = 127 μm sample of teflon. The vertical arrows show the approximate peak energies for black body emission of the room temperature (RT) sample and the quartz-halogen lamp, and the horizontal arrows show the cutoff energy of the long-wave pass filter and an average value for the transmission of teflon, as described in the text. The dashed lines qualitatively indicate how two values of absorption coefficient can be used to approximate the absorption spectrum of thermal radiation.

$$f\, V_{ac} = -A \int dz\, \beta\, e^{-\beta(d-z)}\, \chi\, [-e^{-\alpha z}\, \chi/(\alpha d) + \vartheta(z)/\Psi\,]\, /\, [1-2i\, (\chi/\alpha d)^2] \quad (2a)$$
$$\vartheta(z) \equiv \{[\cosh \chi_{dz} \cos\chi_{dz}) + i \sinh \chi_{dz} \sin \chi_{dz}] - e^{-\alpha d}[\cosh(\chi_z)\cos(\chi_z) + i \sinh(\chi_z) \sin(\chi_z)]\} \quad (2b)$$
$$\text{where}\quad \chi_z \equiv (z/d)\chi \quad \text{and} \quad \chi_{dz} \equiv \chi - \chi_z \quad (2c)$$

(Eq. 2 neglects the effects of internal reflections in the sample.) The integral can be evaluated explicitly and the resulting expression is unchanged if $\alpha$ and $\beta$ are exchanged. ($\alpha,\beta$ exchange equivalence was previously found for the ratio of signals when light illuminated the front and back surfaces.[2]) The calculated in-phase and quadrature responses for a few choices of $\alpha d$ and $\beta d$ are shown in Figure 4 (with the $\alpha=\beta=\infty$ result of Eqtn. (1) also shown for comparison). Note that the fitted value of $f_2$, e.g. corresponding to the peak in $fV_X$ and the step in $fV_Y$, is not very sensitive to the values of $\alpha$ and $\beta$, varying only by ~ 25% between large and small values of the absorption coefficients.

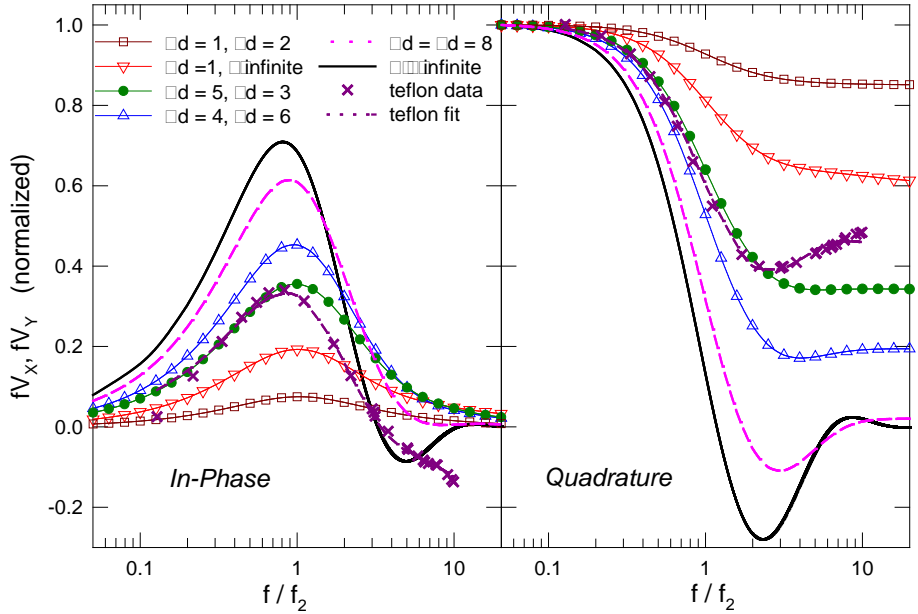

**Figure 4.** (color online) Theoretical frequency dependences of the in-phase ($fV_X$) and quadrature ($fV_Y$) thermal emission signals from a sample for different choices of absorption coefficients $\alpha$ and $\beta$, calculated from Eq. 2. Each curve is normalized to the quadrature signal at low frequency. Also shown is the experimental results for the teflon sample, with $f_2 = 12.3$ Hz, and the fit to the teflon data (with two values of $\beta$, as discussed in the text).

Most importantly, for small $\alpha d$ and/or $\beta d$, $fV_Y$ does not go to zero for frequencies above $f_2$ but there is a shelf in the quadrature response extending to high frequencies. In Ref. [6], we mistook this shelf for the low frequency ($f < f_2$) quadrature response that occurs for large absorption coefficients (see Figure 1). There was also a very large in-phase signal from leaked light, so that we could not fit the in-phase signal and, consequently, a small error in setting the lock-in phase created a drop in quadrature signal at a high frequency, which we mistook for $f_2$. (Similarly, in earlier work we measured the oscillating temperature ($T_{ac}$) on the back surface with a thermocouple glued to the surface. The shelf in $fT_{ac}$ caused by small $\alpha d$ was mistaken for the response expected for $f<f_2$, and in that case the signal dropped at high frequency because of the thermal resistance of the glue holding the thermometer.[5])

If either $\alpha$ or $\beta$ is infinite (in practice, larger than $10/d$), Eq. (2) can be simplified to:[2]

$$fV_{ac} = f(V_X + iV_Y) = -A\chi[(B_X + C_X + D_X) + i(B_Y + C_Y + D_Y)]/[1 + 4(\chi/\gamma d)^4] \quad (3a)$$
$$B_X = \{-[1 + 2(\chi/\gamma d)^2]\sinh\chi\cos\chi + [1 - 2(\chi/\gamma d)^2]\cosh\chi\sin\chi\}/|\Psi|^2 \quad (3b)$$
$$B_Y = \{[1 - 2(\chi/\gamma d)^2]\sinh\chi\cos\chi + [1 + 2(\chi/\gamma d)^2]\cosh\chi\sin\chi\}/|\Psi|^2 \quad (3c)$$
$$C_X = \exp(-\gamma d)\{[1 + 2(\chi/\gamma d)^2]\cosh\chi\sinh\chi - [1 - 2(\chi/\gamma d)^2]\sin\chi\cos\chi\}/|\Psi|^2 \quad (3d)$$
$$C_Y = -\exp(-\gamma d)\{[1 - 2(\chi/\gamma d)^2]\cosh\chi\sinh\chi + [1 + 2(\chi/\gamma d)^2]\sin\chi\cos\chi\}/|\Psi|^2 \quad (3e)$$
$$D_X = (\chi/\gamma d)\exp(-\gamma d), \quad D_Y = 2((\chi/\gamma d)^3\exp(-\gamma d) \quad (3f)$$

Here $\gamma$ is whichever of $\alpha$ or $\beta$ is finite. (Also, if both $\alpha$ and $\beta$ are finite, one can *approximately* replace them, for frequencies within a few times $f_2$, with a single effective $\gamma$, and use Eq. (3).)

We have checked these expressions by measuring the photothermal response of a 127 μm thick piece of teflon (polytetrafluoroethylene), the infrared transmission spectrum of which is shown in Figure 3. Measurements were made with the diode laser and with a ~ 200 nm (visibly opaque) PbS film evaporated on the front surface to increase the absorption (making $\alpha d \gg 10$ and $\gamma \approx \beta$). The data and fits are shown in Figure 4. For the fit, we assumed two values of $\beta$; the resulting fit had $f_2 = (12 \pm 1)$ Hz, with fitted $\beta$ values of $\infty$ (i.e. $\gg 10$, corresponding to the opaque regions of the spectrum for $\nu < 1000$ cm$^{-1}$)) and $\approx 0.8$, i.e. corresponding to the average transmission value shown by the horizontal arrow in Figure 3. From the value of $f_2$, we find $D = (0.13 \pm 0.01)$ mm$^2$/s, consistent with the measured value near room temperature.[15]

We have remeasured the photothermal response for crystals of TIPS-pentacene with thicknesses ranging from 50 μm to 270 μm. Crystals with areas > 3 mm$^2$ generally do not have uniform thicknesses, but may be wedge shaped or have stepped surfaces, and the resulting uncertainties in the sample thicknesses for the crystals we measured range from ±10 to ±20 μm. To avoid having a large range of incident absorption coefficients that would result from the incandescent source, we used the blue laser as the light source (making $\alpha \gg \beta$). The data and their fits to Eq. (3) are shown in Figure 5. We included two values of $\gamma$ ($\approx \beta$, corresponding to the dashed lines shown in Figure 4) as fitting parameters. (The fits tend to overestimate the in-phase response at low frequencies, presumably because the frequency is beginning to approach $1/(2\pi\tau_1)$.)

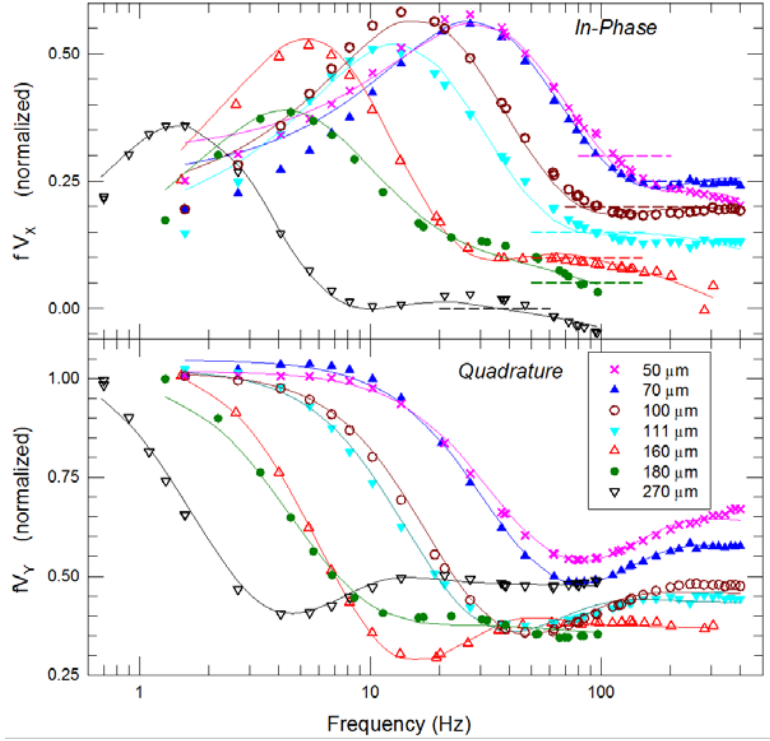

**Figure 5**. (color online) In-phase and quadrature responses for TIPS-pentacene crystals of different thicknesses, as indicated.; the solid curves show the fits to Eq. (3) with two values of $\gamma$. The signals were (approximately) normalized to the quadrature response of each at low frequency. Subsequent in-phase graphs are vertically offset by 0.05 and the dashed lines show the zero-lines for each graph. (Note that the in-phase and quadrature responses are plotted with different vertical scales.)

The variation of the fitted values of $\tau_2 = 1/(2\pi f_2)$ with $d^2$ is shown in Figure 6. From the slope we find that the interlayer (c axis) diffusivity $D = (0.10 \pm 0.01)$ mm$^2$/s. This value is two orders of magnitude smaller than estimated in Ref. [6] and corresponds to an interlayer thermal conductivity $\kappa_c = 0.17$ W/m·K $\approx \kappa_{ab}/10$, where $\kappa_{ab}$ is the thermal conductivity in the in-plane, needle-axis, high electronic mobility direction.[5] This value of $\kappa_c$ is 65% larger than the value found for sublimed films,[14] and the difference can be readily associated with disorder in the films: from the analysis of Ref. [14], if one assumes that the heat is carried by acoustic phonons only, the value of $\kappa_c$ implies a phonon mean-free path of 3-4 c (the interlayer spacing) in the crystal and ~ 2c in the film. The value of $\kappa_c$ in the crystal is also about twice that found for crystals of rubrene,[9] another layered small molecule organic semiconductor.

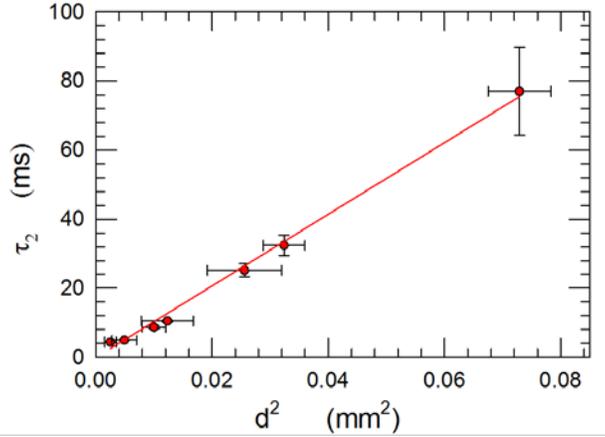

**Figure 6.** (color online) The thickness dependence of fitted values of $\tau_2 = 1/(2\pi f_2)$ for crystals of TIPS-pentacene. The slope $a = (1040 \pm 100)$ ms/mm$^2$ determines the interlayer diffusivity $D = 0.105 / a = (0.10 \pm 0.01)$ mm$^2$/s.

## IV. Conclusion

The ac-photothermal technique described above represents a relatively inexpensive technique to measure the through-plane thermal diffusivities of thin samples, e.g. d < 0.5 mm, using readily available equipment. It is especially useful for new materials for which available samples are either too thin (e.g conducting polymers) or have too small a surface area (e.g. organic crystals) for measurements with a light-flash apparatus. The analysis is very straight-forward for opaque materials (i.e. Eq. (1)) but can be extended to non-opaque samples using Eq. (3). In either case, we have typically found excellent agreement with the theoretical equations for frequencies within a decade of the characteristic frequency, $f_2$.

Use of Eqtns. (2,3) requires some care. If the sample is not opaque (e.g. $\alpha d < 10$) to incident radiation, one should use a monochromatic light source and not an incandescent light source, for which there will presumably be a wide distribution of $\alpha$-values. This is typically not a problem for the emitted thermal radiation, since $\beta$ can generally be approximated by one or a few "average" values (as we did for TIPS-pentacene), especially if an LPF is used to limit the spectral range of detected radiation. In this case, one can use Eq. (3) in the calculation (and, if needed, fitting to an effective value of $\gamma$ that combines the effects of finite $\alpha$ and $\beta$). Uniform evaporated or sputtered films can also be used to improve the optical properties of samples, but one should be sure that they stick well and have small interface thermal resistances.

While our present experiments were limited to frequencies below 2 kHz, this is not a fundamental limitation. Use of a higher frequency chopper or high bandwidth modulated laser would allow measurements of thinner samples. For example, measurements at 200 kHz would allow measurements of PEDOT:PSS samples as thin as 1 μm.

As an example of a non-opaque material, we restudied crystals of TIPS-pentacene, for crystals varying in thickness from 50 to 270 μm. The resulting value of the interlayer thermal diffusivity, $D \approx 0.10$ mm$^2$/s, is much smaller than that we previously reported when we overlooked the effects of finite absorption length,[6] and is consistent with thin film measurements[14] and values expected for a layered organic crystal; for example, this value is much smaller than the needle-axis diffusivity,[5] as expected.

We thank Z. Li, J. Edberg, F. Zhang, and X. Crispin (Linkoping U.) for providing samples of NFC-PEDOT and PEDOT:PSS, M.M. Payne and J.E. Anthony (U. Kentucky) for both providing samples of TIPS-pentacene and helpful suggestions in preparing new crystals, and Y. Yao and M. Weisenberger (U. Kentucky) for helpful discussions on techniques. This research was supported by the National Science Foundation, Grant No. DMR-1262261.